# Efficient Screening of Organic Singlet Fission Molecules Using Graph Neural Networks

Li Fu[1], Longfei Lv[1], Fan Zhang[2] Si Zhou[1], Weiwei Gao[3, 4*], Jijun Zhao[1*]

[1] *Guangdong Basic Research Center of Excellence for Structure and Fundamental Interactions of Matter, Guangdong Provincial Key Laboratory of Quantum Engineering and Quantum Materials, School of Physics, South China Normal University, Guangzhou 510006, China*
[2] *School of Physics, Nanjing University, Nanjing 210093, China*
[3] *Institute of Atom Manufacturing, Nanjing University, Suzhou 215163, China*
[4] *Nanjing Institute of Atomic Scale Manufacturing, Nanjing 211800, China*

## Abstract

Singlet fission (SF) provides a promising strategy for surpassing the Shockley-Queisser limit in photovoltaics. However, the identification of efficient SF materials is hindered by the limited availability of suitable molecular candidates and the high computational costs associated with conventional quantum-chemical methods for excited states. In this study, we introduce a high-throughput screening framework that integrates a graph neural network (GNN) with multi-level validation to accelerate the discovery of SF-active molecules. Trained on a previously reported FORMED database, the GNN achieves state-of-the-art accuracy in predicting SF-relevant excited-state properties, demonstrating a mean absolute error of about 0.1 eV for $S_1$, $T_1$, and $T_2$ excitation energies. This capability facilitates the efficient screening of over 20 million molecular structures from both OE62 and QO2Mol databases. Our framework significantly reduces the computational demand associated with Time-Dependent Density Functional Theory validation by four orders of magnitude and identifies 180 potential SF molecules along with more than 1000 conformers. Subsequent assessments regarding synthetic accessibility, GW approximation and Bethe-Salpeter equation calculations further highlight a subset of experimentally feasible candidates among these SF candidates. The approach presented herein exemplifies an effective strategy for accelerating the discovery of functional molecules with optoelectronic applications.

---
*Corresponding authors. *E-mail*: zhaojj@scnu.edu.cn (Jijun Zhao); weiweigao@nju.edu.cn (Weiwei Gao)

# 1. Introduction

Singlet fission (SF) is a photophysical phenomenon observed in organic systems, wherein a high-energy singlet exciton ($S_1$) undergoes a process of splitting into two lower-energy triplet excitons ($T_1$). This mechanism has the potential to effectively double the photocurrent generated from a single high-energy photon. [1-8] Therefore, it plays an important role in photovoltaic devices by capturing excess energy from high-energy photons and thereby reducing thermalization losses. Molecules capable of undergoing SF have the potential to enhance the power conversion efficiency of solar cells beyond the Shockley-Queisser limit, while the reduction of thermalization losses may also lower module temperatures and extend device lifetimes.[9] Since its first invocation in 1965 to elucidate the photophysics of anthracene crystals, [10] the phenomenon of SF has been reported in acene derivatives, benzofurans, carotenoids, and conjugated polymers through a series of experiments. [1,3] However, most of these candidates exhibit inadequate ambient stability, while the overall count of verified SF molecules remains relatively low. This shortage of appropriate SF materials not only obstructs the commercialization efforts for SF-based solar cells but also impedes a comprehensive understanding of the underlying mechanisms governing singlet fission. Considering the vast chemical diversity of organic compounds, numerous potential candidates for singlet fission have yet to be investigated, thereby rendering the identification and screening process for novel SF-active molecules a significant challenge within this research domain.

On the other hand, the theoretical identification of SF molecules remains a non-trivial task, primarily due to the need to satisfy multiple criteria. A fundamental requirement is that the SF process must be thermodynamically favorable; specifically, this entails that the energy of the singlet excited state exceeds twice that of the triplet excited state ($E_{S1}-2E_{T1} \geq 0$). This energy criterion has been widely recognized as a driving force for the SF process in previous studies [11]. Another important condition is that the higher-lying triplet state maintains a sufficient energy gap relative to the first triplet ($E_{T2}-2E_{T1} \geq 0$) in order to prevent triplet-triplet annihilation [12]. Furthermore, the SF material system used for photoelectric conversion should exhibit strong absorption in the visible light region. To effectively harvest triplet excitons, it is essential that $E_{T1} \geq E_g$ (for instance, considering commonly used silicon photovoltaic materials where $E_g$ is approximately 1.1 eV); thus, its triplet energy level should exceed the band gap of the interfaced material. In addition to these energy-level criteria, practical SF-active molecules must possess high chemical stability and synthetic accessibility to be suitable for experimental and practical applications.

The theoretical screening of SF molecules necessitates the assessment of their singlet and triplet excitation energies, followed by identification based on the aforementioned energy level screening criteria. Among various first-principles approaches, time-dependent density functional theory (TDDFT) provides a reasonable balance between computational efficiency and accuracy in excited-state calculations. Consequently, it is widely utilized for evaluating molecular excitation energies. For example, Padula *et al*. [13] performed TDDFT calculations to screen 40,000 molecules with no more than 100 atoms from the Cambridge Structural Database and identified over 200 candidates that satisfy the energy criterion of $E_{S1} \geq 2E_{T1}$. Similarly, Perkinson *et al*. [14] carried out high-throughput TDDFT calculations to perform virtual screening of 4,482 organic molecules with anthracene substructures, which were retrieved from the eMolecules and Reaxys databases. They identified 88 organic molecules with potential SF properties; among them, two molecules were successfully synthesized in subsequent experiments. Beyond TDDFT approaches, excitation properties can also be described using many-body perturbation theory (MBPT) within the GW approximation combined with the Bethe-Salpeter equation (GW+BSE). Previous studies have shown that the Tamm–Dancoff approximation (TDA) [15] effectively mitigates triplet instabilities in GW+BSE calculations, improving the accuracy of both singlet and triplet energies for gas-phase organic molecules. [16,17]. Liu *et al*. [17] applied MBPT-based thermodynamic screening combined with a SISSO algorithm [18] to evaluate 101 polycyclic aromatic hydrocarbons and identified three promising SF candidates. While GW+BSE typically offers enhanced accuracy in the characterization of excited states, its computational expense renders it impractical for high-throughput screening or large-scale calculations. The trade-off between accuracy and scalability highlights the need for more efficient yet reliable computational screening strategies.

As an emerging paradigm of materials discovery, machine learning (ML) techniques have recently been utilized to predict and expedite the identification of SF molecules. Zhu *et al*. [19] proposed Catalyst Deep Neural Networks to predict the SF properties of anthracene-based molecules reported by Perkinson *et al*.[14], achieving an impressive prediction accuracy of 98%. Based on support vector machine algorithm, Borislavov *et al*. [20] also developed a binary classification model to screen general-purpose data sets for potential SF candidates according to diradical character of the molecules. Most recently, Corminboeuf *et al*. [12] constructed the fragment-oriented materials design database (FORMED) and trained an XGBoost model to quickly predict molecular properties such as band gap and excitation energy, which has a prediction error of approximately 0.2 eV for excited state properties. Furthermore, they integrated an automated assembly method with an uncertainty-controlled genetic algorithm to investigate previously inaccessible regions of the organic chemical space. This innovative approach resulted in the identification of 95 top candidates, among which 8 met the adiabatic SF criterion. [21]

Although previous studies have made strides in accelerating the screening process for SF molecules, two significant challenges remain. Firstly, current ML models often exhibit limited accuracy and poor transferability, primarily due to the absence of large and diverse datasets that encompass excited-state properties. Secondly, most existing screening efforts rely on a singular criterion (e.g., driving force), which fails to provide a comprehensive evaluation essential for identifying experimentally viable candidates. To address these issues, we developed a graph neural network (GNN) trained on the FORMED database, achieving a mean absolute error of approximately 0.1 eV in energy predictions of exciton states. This GNN model facilitates the rapid estimation of excited-state properties for over 20 million organic molecules within the OE62 and QO2Mol datasets. By integrating GNN predictions with a limited number of TDDFT calculations, we streamlined the screening workflow and identified thousands of promising SF candidates. Impressively, the computational cost associated with TDDFT calculations was reduced by four orders of magnitude. Molecules exhibiting high synthetic feasibility—independently filtered by the DeepSA model—were further prioritized for validation through GW-BSE computations. These candidates hold substantial potential to mitigate the current scarcity of suitable materials for practical SF-based photovoltaic applications.

## 2. Strategy

An efficient framework that combines a GNN with a singlet fission scoring function is proposed for the high-throughput screening of SF molecules (**Figure 1**). Our screening workflow begins with the recently developed FORMED dataset, which contains over 110,000 organic molecules and was utilized to train the GNN model for predicting properties such as the HOMO-LUMO gap ($E_{HL}$) and excitation energies ($E_{S1}$, $E_{T1}$, $E_{T2}$). Leveraging this well-trained GNN model, we predicted excited-state properties for more than 20 million organic molecules across two datasets: OE62 and QO2Mol, based on their geometric structures. Candidates that met the SF energy criteria were subsequently validated using TDDFT calculations. A comprehensive scoring function that incorporates all relevant energetic conditions was then applied to quantify the potential of these molecules for singlet fission. Another critical aspect of molecular design is synthetic accessibility; thus, we evaluated the ease of synthesis for SF candidates filtered through TDDFT calculations. The 79 molecules identified by the DeepSA model that possess high synthetic accessibility were further subjected to GW+BSE calculations for validation.

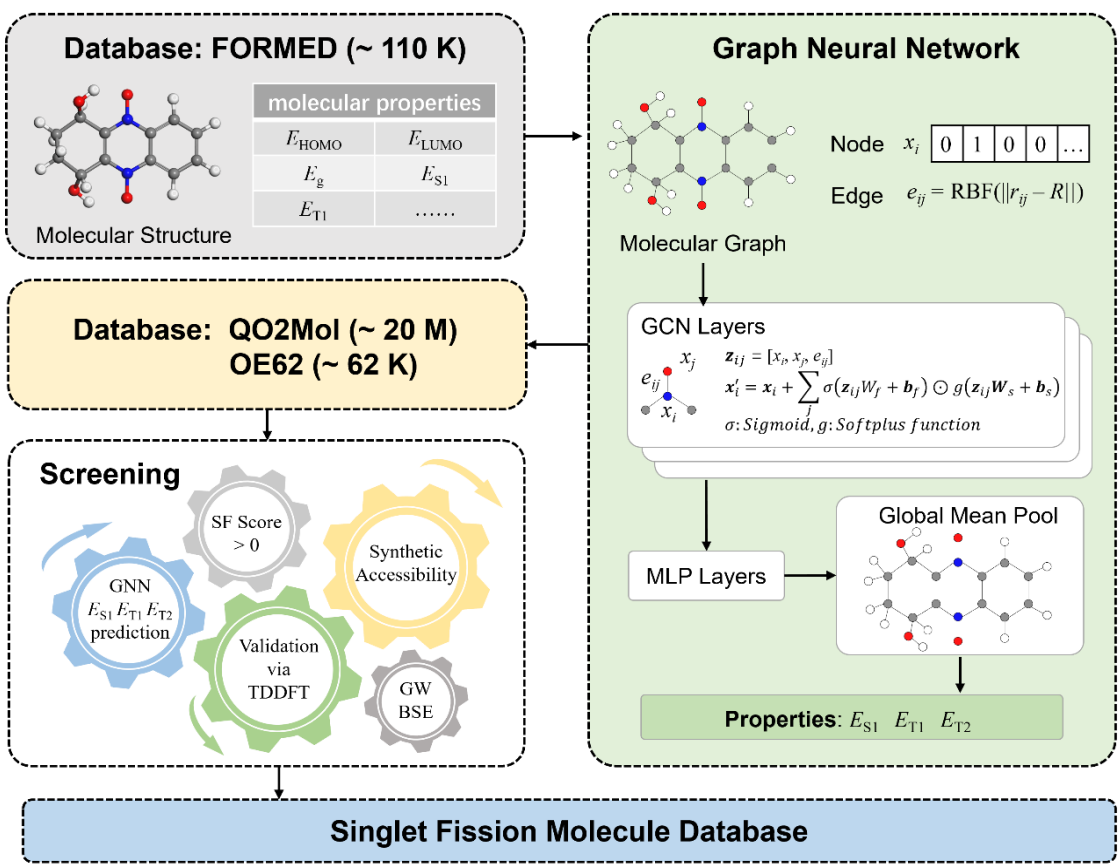

**Figure 1**. Screening workflow for singlet fission molecules.

## 2.1 Datasets

The GNN model developed for predicting excited-state properties was trained using the FORMED database. The FORMED dataset was meticulously curated from experimental databases through processes such as element selection, isomer identification, and connectivity verification during optimization [12]. It comprises over 110,000 organic molecular structures along with their associated properties. In comparison to the widely used QM9 database [22], the FORMED database encompasses larger molecules and a more diverse array of elements; in addition to C, N, O, F, and H, it also includes elements such as B, Si, P, S, Cl, As, Se, and Br. The molecular sizes within this dataset span a broad range, from small molecules containing ten atoms to extensive molecular systems comprising over 200 atoms. All molecules in the FORMED database have undergone geometric optimization using the GFN2-xTB semiempirical method [23], followed by DFT and TDA-TDDFT calculations to obtain ground-state energies and excitation energies at the ωB97X-D/6-31G(d) level. Specifically, this comprehensive database provides 116,687 stable geometric structures of organic molecules alongside their corresponding ground-state and excited-state energies, information of molecular orbitals (i.e., HOMO, LUMO, HOMO-LUMO Gap), the first five singlet excitation energies, the first five triplet excitation energies, as well as key excited-state parameters such as oscillator strengths. The extensive collection of excited-state data contained within the FORMED database offers a robust and reliable foundation for machine learning models aimed at predicting these properties effectively.

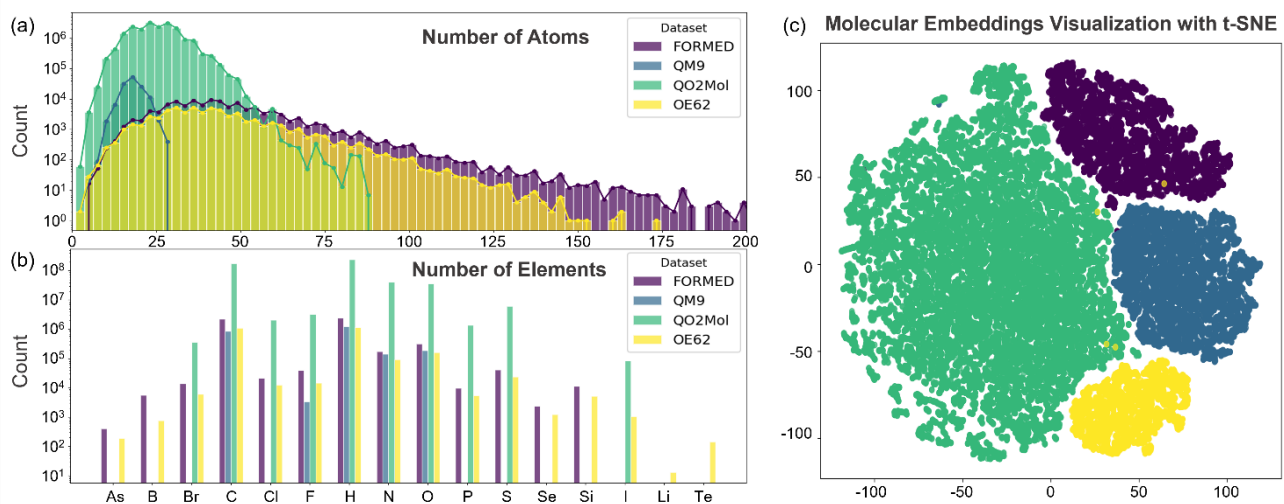

**Figure 2**. Chemical space spanned by FORMED, QM9, OE62 and QO2Mol dataset. (a) Molecular size distributions (including hydrogen atoms). (b) Distribution of different element types in four datasets. (c) *t*-distributed stochastic neighbor embedding plot generated from GNN model.

With the ongoing advancement of high-throughput experimental and computational techniques, a multitude of databases housing extensive collections of organic molecular structures have been established. For example, the OE62 database, compiled by Stuke *et al.* [24], consists of 61,489 distinct organic molecules extracted from 64,725 experimental crystal structures collected across various application fields by Schober *et al.* [25]. All structures have been fully optimized using DFT calculations at the PBE level with Tier2 basis sets. OE62 also encompasses the largest variety of elements among the explored datasets, i.e.,16 different elements in total, with molecular sizes ranging from a few atoms to over one hundred atoms (Figure 2a, b). On the other hand, Liu *et al.* [26] generated over 12,000 fragments and more than 20 million conformations of organic molecules from ChEMBL structure. All these molecular structures were then optimized by DFT calculations at the B3LYP/def2-SVP level and collected in the QO2Mol database. Molecules in QO2Mol span ten elements (C, H, O, N, S, P, F, Cl, Br, and I), with heavy-atom counts exceeding 40. In addition, the QM9 dataset, frequently utilized as a benchmark for machine learning of molecules, is also presented in **Figure 2** for comparison. It contains 133,885 small molecules with no more than nine heavy atoms, i.e., C, H, O, N, and F. To visualize the chemical space covered by these datasets, *t*-distributed Stochastic Neighbor Embedding (*t*-SNE) algorithm [27] was utilized to reduce the high-dimensional molecular features learned by a GNN (specifically, a GNN with two convolutional layers was trained for ten epochs). As shown in **Figure 2c**, the four datasets occupy distinct regions in the two-dimensional space, reflecting their molecular diversity and supporting the rationality of the current strategy for molecular graph construction.

## 2.2 Singlet Fission Score

As is well known, SF molecules often need to satisfy certain energetic criteria (**Figure 3a**), which can be succinctly summarized as $E_{S1} \geq 2E_{T1}$ and $E_{T2} \geq 2E_{T1}$. Furthermore, the energy of the triplet excited state of the molecule must also meet the condition $E_{T1} \geq E_g$ (where $E_g$ = 1.1 eV for crystalline silicon material). To quantitatively evaluate SF potential, we have defined a score function $\xi$, which is described in **Figure 3b**. A molecule that meets all energy level criteria is assigned a score of $\xi \geq 0$; otherwise, $\xi < 0$, with molecules that are far from the target region receiving increasingly negative scores. We applied this scoring function to all molecules in the FORMED database, and their distribution in three-dimensional space is illustrated in **Figure 3**.

Noticeably, only a limited number of molecules achieve a positive score in the FORMED database. For the majority of molecules, the difference between the first singlet excited-state energy and the first triplet energy is smaller than the triplet energy itself. Moreover, the $E_{T1}$ values of these candidates are concentrated within a relatively narrow range, approximately 2 to 4 eV. Furthermore, correlation analysis of the FORMED database revealed that molecular planarity exhibits an extremely weak negative correlation with SF propensity, whereas aromaticity and the number of six-membered rings display a weak positive correlation with SF scores (**Figure S1**). In other words, only qualitative insights can be drawn from the geometric features of these molecules, which are insufficient for accurately predicting the SF properties of unknown compounds. This underscores the necessity for more advanced machine learning approaches—such as the Graph Neural Network (GNN) model discussed in the following section—to provide a quantitative description.

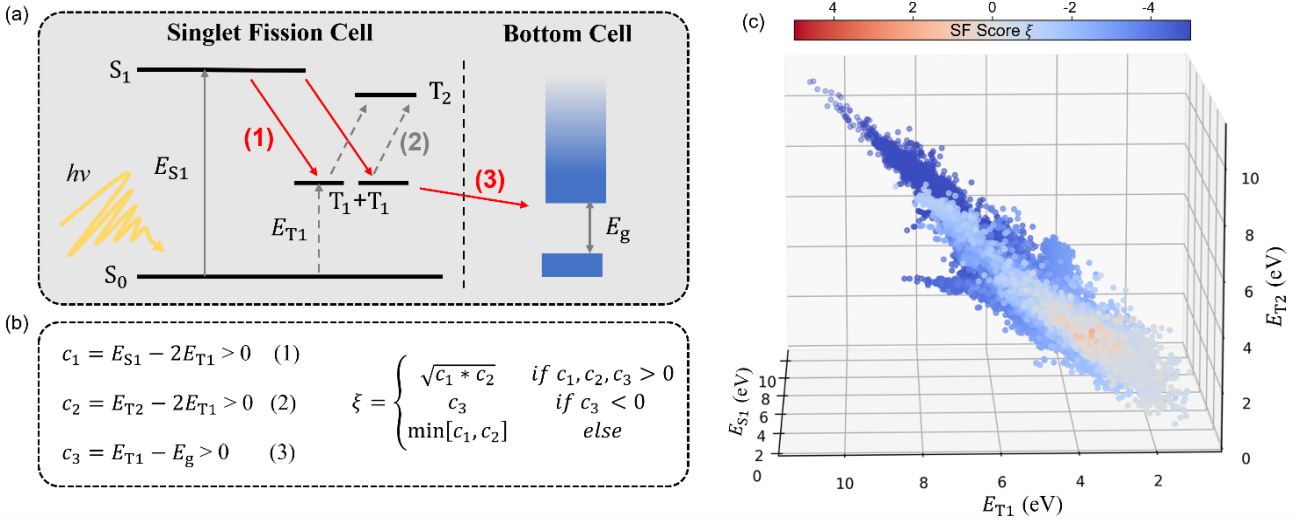

**Figure 3**. SF scores of molecules in the FORMED database. (a) SF process. (b) SF score. (c) Excitation energies and their SF scores, whereas red (blue) color corresponds to a SF molecule with positive (negative) SF score.

## 2.3 Graph Neural Network

The structure of the GNN model employed in this work is illustrated in **Figure 1**. Each molecule, represented by its atomic coordinates, is first converted into a molecular graph with a cutoff radius of $R_C$ = 5 Å, which is encoded into an adjacency matrix:

$$A_{ij} = \begin{cases} 1 & ||r_i - r_j|| \leq R_c \\ 0 & ||r_i - r_j|| > R_c \end{cases}$$

The node features in the molecular graph are represented using the 92-dimensional vector $x_i$ from the crystal graph convolutional neural network (CGCNN), which includes elemental properties such as electronegativity, covalent radius, atomic orbitals, and valence electrons. [28] The initial features of edge $e_{ij}$ are the expansion coefficient vectors of radial basis functions of the interatomic distances, with each edge corresponding to a 20-dimensional vector. The main body of the network consists of multiple graph convolution layers (GCN) and fully connected layers. The node features after the convolution operation are updated using the following formula:

$$x'_i = x_i + \sum_j \sigma(z_{i,j} W_f + b_f) \odot g(z_{i,j} W_s + b_s)$$

Here, feature matrix $z_{i,j}$ is obtained by concatenating $x_i$, $x_j$, and the edge feature $e_{ij}$ between the two atoms; $\sigma$ and $g$ represent the sigmoid and softplus activation functions, respectively; $\odot$ denotes the Hadamard product. After multiple convolution operations to aggregate the features of neighboring nodes and their own features, each node's feature is mapped to a specified dimension through multiple fully connected layers. Then, a global mean pooling layer is applied to the features of all nodes within the graph, resulting in the final output representing the value of the property predicted by the GNN model. During the training process, the Adam method [29] was used with a learning rate of 0.001~0.0005, and the loss function is the mean absolute error (MAE). After thorough hyperparameter optimization (**Figure S2**), our GNN model demonstrated satisfactory performance in predicting the excited-state properties of the FORMED database. Both the construction of molecular graphs and the implementation of GNN were conducted utilizing the PyTorch [30] framework along with the PyTorch-Geometric package.[31]

## 2.4 TDDFT and GW+BSE

To ensure data consistency, we conducted the validation of prediction results from graph neural networks using TDDFT by calculating the excited-state properties of the molecules with the same computational settings used for the FORMED dataset. The vertical excitations were computed with TDA-TDDFT based on Gaussian16 package at the ωB97X-D/6-31G(d) level. The ground state electronic properties of selected molecules with potential SF character were independently computed with the QUANTUM ESPRESSO [32] based on density functional theory (DFT) and plane-wave basis. The Perdew-Burke-Ernzerhof (PBE) functional was used to account for the exchange-correlation interaction. The optimized norm-conserving pseudopotentials were adopted for describing the ion-electron interaction. [33,34] The Kohn–Sham orbitals were expanded in plane waves with a kinetic energy cutoff of 90 Ry, and the Brillouin zone was sampled by the Γ point. The calculations of excitonic properties for molecule are carried out by solving the BSE within the TDA, [35,36] as implemented in the BERKELEYGW package [37]. We computed the dielectric matrices with a polarizability cutoff of 12 Ry. Spin-triplet and spin-singlet excitations were calculated separately by diagonalizing BSE by setting $K_{single}^{eh} = K_x + K_d$ and $K_{tiplet} = K_d$, where $K_{eh}$ is the electron-hole interaction kernel, $K_x$ and $K_d$ are exchange and direct interaction terms, respectively. The matrix elements of the BSE Hamiltonian were explicitly calculated using 10 valence and 10 conduction bands.

## 3. Results and Discussion

## 3.1 Train and test of GNN

To predict the properties of molecules, we developed a GNN model, which is illustrated in **Figure 1**. In GNN model, molecular structures are abstracted into molecular graphs, where atomic element types define the initial node features and the distances between atoms are used as edge features. Information from each atom and its neighbors is progressively aggregated through several graph convolution layers, and the resulting representation is converted into predicted molecular properties via a global mean pool layer and fully connected layers. The FORMED database [12] was randomly divided into train and test sets with an 8:2 ratio to train and test the network. Taking the task of HOMO-LUMO gap prediction as an example, **Figure S2** illustrates the impact of the number of GCN layers and the number of nodes in the fully connected layers on the overall performance of network. When the number of GCN layers exceeded nine, the test error no longer decreases appreciably. Increasing the number of hidden nodes in the fully connected layers also brings about negligible improvement. Overall, a 10-layer GCN with 64 hidden nodes achieves the best performance in band gap prediction, yielding a MAE of only 0.16 eV. The present model for gap prediction substantially prevails the previous reports, e.g., MAE of 0.26 eV by the XGBoost model [12] and MAE of 0.30 eV by SchNet model [38]. The data points from the test set predicted by the GNN are compared with the TDDFT results in **Figure 4a**, exhibiting a close alignment along the diagonal and a coefficient of determination ($R^2$) greater than 0.96. This strong correlation further underscores the high predictive accuracy of our GNN model.

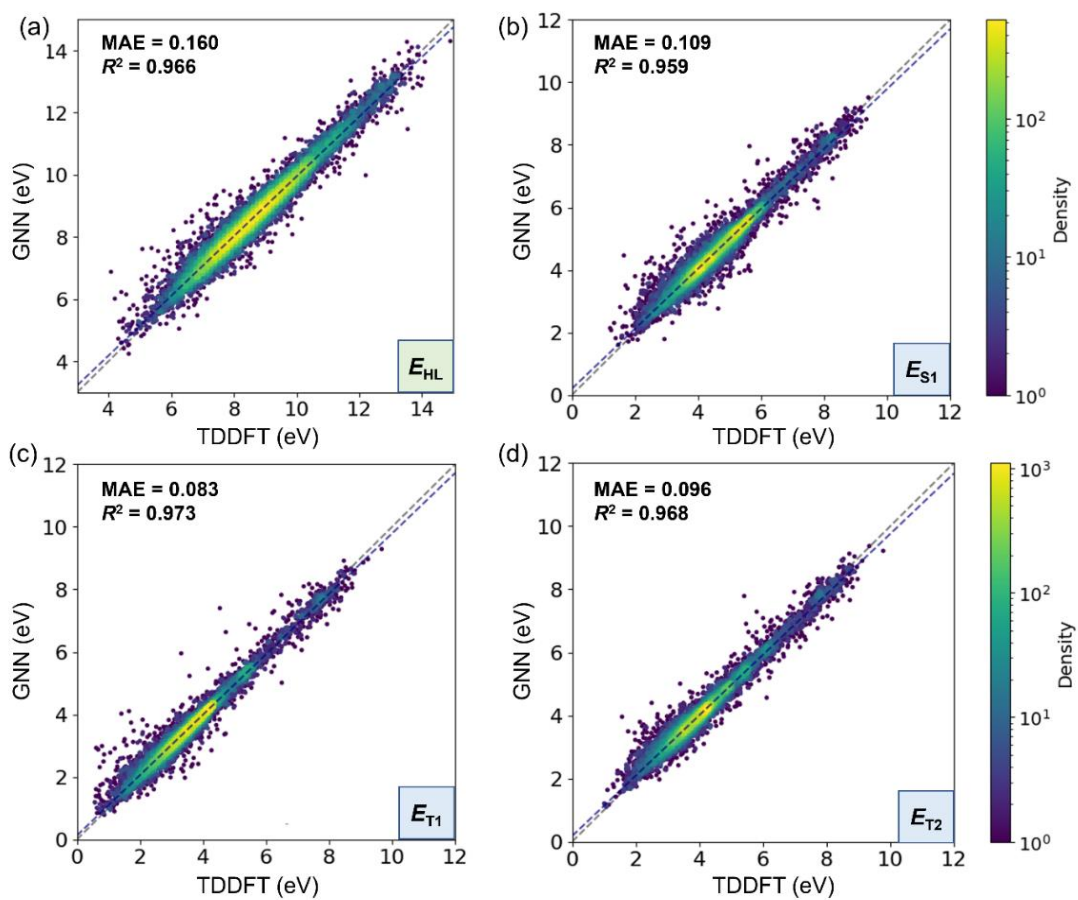

**Figure 4**. Comparison of HOMO-LUMO gap ($E_{HL}$) and excitation energies ($E_{S1}$, $E_{T1}$, $E_{T2}$) between GNN prediction and TDDFT calculation for the test subset of the FORMED dataset.

The same network architecture was also employed to train models for excited-state properties, including the excitation energies of S$_1$, T$_1$, and T$_2$ states (see **Figure 4b, c, d**). The GNN model demonstrates excellent performance in predicting these excited-state properties, achieving a test-set MAE of approximately 0.1 eV and $R^2$ exceeding 0.95 for all states. Compared with the XGBoost model, [12] which yields prediction errors of 0.20 eV for $E_{S1}$ and 0.18 eV for $E_{T1}$, respectively, the GNN gains nearly a 50% improvement in accuracy. Furthermore, we evaluated the performance of a classical graph neural network model, namely, SchNet [38] on the FORMED database. All test errors are summarized in **Table 1**, showing the best performance of our GNN model.

Table 1. Test Errors of properties prediction in FORMED dataset using GNN, SchNet and XGBoost models.

|  | XGBoost | SchNet | GNN (this work) |
| --- | --- | --- | --- |
| MAE of $E_{HL}$ (eV) | 0.26 | 0.30 | 0.16 |
| MAE of $E_{S1}$ (eV) | 0.20 | 0.20 | 0.11 |
| MAE of $E_{T1}$ (eV) | 0.18 | 0.17 | 0.08 |

## 3.2 Prediction

Molecular structures from the OE62 and QO2Mol databases were transformed into molecular graphs and then sent to the well-trained GNN model for excitation energy prediction. This model enables the rapid estimation of $E_{S1}$, $E_{T1}$, $E_{T2}$ for over 20 million molecules, thereby allowing efficient screening of potential SF molecules according to the required energy condition. Notably, considering the inherent prediction error of approximately 0.1 eV in the GNN results, the stringent energy-matching criteria should be slightly relaxed to avoid the exclusion of promising singlet SF candidates. To this end, we introduced a cutoff energy $\delta$, which relaxes each condition on the equation as follows:

$$c_1, c_2, c_3 \geq -\delta$$

where $c_1$, $c_2$, $c_3$ is defined in **Figure 3b**. When $\delta = 0$ eV, the condition corresponds to a strict energy-level matching criterion. Using the GNN model, the numbers of SF molecules predicted from the screening procedure under different cutoff energies (0, 0.2, 0.5, and 1.0 eV) across the two databases are summarized in **Table 2**. In both datasets, loosing the cutoff energy threshold leads to a pronounced increase in the number of candidate SF molecules. Therefore, to ensure the reliability of the identified candidates, further computational validation through TDDFT calculations is indispensable.

**Table 2**. The numbers of potential SF molecules screened from QO2Mol and OE62 databases under different cutoff energy ($\delta$) based on their excitation energy predicted by GNN. The values in parentheses are the numbers of SF molecules validated by TDDFT calculations.

| $\delta$ (eV) | OE62 | QO2Mol |
| --- | --- | --- |
| 0 | 107 (64) | 2912 (566) |
| 0.2 | 237 (88) | 5560 (1080) |
| 0.5 | 553 (105) | 16919 |
| 1 | 1825 (118) | 97244 |

Taking the OE62 dataset as an example, we performed TDDFT calculations on all 1,825 candidate molecules identified under an energy cutoff of 1.0 eV. For consistency, TDDFT calculations were carried out using Gaussian16 at the ωB97X-D/6-31G(d) level of theory. The comparison between the TDDFT-computed $E_{S1}$, $E_{T1}$ and $E_{T2}$ values and those predicted by the GNN model is shown in **Figure 5**. Notably, the GNN model—trained solely on the FORMED database—achieves a mean error of merely 0.13 eV when evaluated on the completely independent OE62 dataset, demonstrating its excellent transferability and consistency. Although the screened molecules are primarily concentrated in the lower excitation-energy region after applying the energy-level matching criteria, the $R^2$ remains sufficiently high, indicating robust predictive performance.

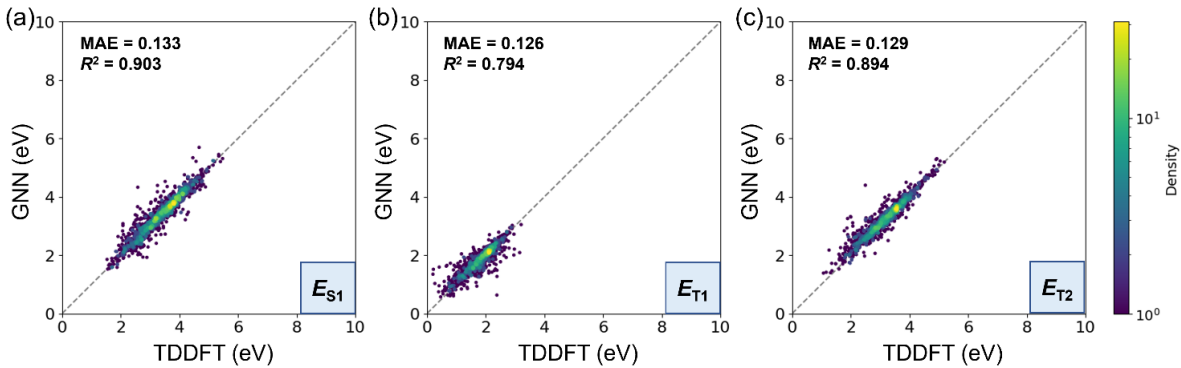

**Figure 5**. Excitation energies of candidate molecules screened by the OE62 database at $\delta$ = 1 eV predicted by GNN model and calculated by TDDFT.

As given in **Table 2**, the candidate molecules corresponding to different cutoff energies exhibit a hierarchical inclusion relationship. Following TDDFT validation and the application of strict energy-level matching criteria, expanding the candidate pool by loosening the energy thresholds did not lead to a significant increase in the number of SF molecules identified in the OE62 dataset. This finding suggests that the GNN model achieves high accuracy in predicting the excited-state properties. Specifically, a total of 1,825 candidates were selected from the OE62 dataset using GNN model with $\delta = 1$ eV, whereas 118 of them were further identified by TDDFT calculations. To screen the QO2Mol database, which contains over 20 million molecules, it is essential to strike a careful balance between the breadth of screening and computational cost. To determine an appropriate value of $\delta$ for the QO2Mol database, we performed a statistical analysis on the number of TDDFT-validated SF molecules retrieved from the OE62 database using different $\delta$ thresholds. Compared to the dataset of SF candidates identified with $\delta = 1$ eV, 54%, 75%, and 89% were recovered when $\delta$ was set to 0, 0.2, and 0.5 eV, respectively, resulting in computational cost reductions to 6%, 13%, and 31%, respectively. Assuming that $\delta = 1$ eV captures all potential candidates, it can be inferred that performing TDDFT validation on about 5,000 molecules prescreened with $\delta = 0.2$ eV would be sufficed to identify around 75% of the SF molecules in the QO2Mol database. In summary, the proposed strategy effectively reduces the computational burden of TDDFT calculations by roughly four orders of magnitude while retaining a substantial fraction of promising SF candidates.

Among the 5,560 molecules in the QO2Mol database validated TDDFT calculations, 1,080 structures exhibited SF scores ($\xi \geq 0$). By clustering these molecules according to their SMILES strings [39], it was found that these conformers correspond to only 62 distinct molecules. This observation indicates that different geometric conformations of the same molecule—sharing identical formula and atomic connectivity—generally retain their SF properties, despite minor variations such as substituent rotations or slight conformational distortions. Moreover, although our screening workflow and GNN model are based on molecular geometry, they exhibit notable robustness in identifying SF-capable molecules. Ultimately, 118 SF candidates from the OE62 database and 1,080 SF candidates from the QO2Mol database were carefully examined for duplication using their SMILES representations. All molecules sharing identical SMILES are listed in **Table S1** of the Supporting Information.

### 3.3 Singlet Fission Molecules

By screening of the OE62 and QO2Mol databases, we identified 180 organic molecules that possess singlet fission properties at the TDDFT level. The corresponding structure files and low-lying excited-state properties are provided in the Supporting Information. These molecules consist of a diverse array of elements and display a wide range of structural motifs, including derivatives of commonly reported acene-type compounds. In the context of SF molecular design, candidate structure identification through high-throughput screening represents one crucial aspect; an equally important consideration is the feasibility of their experimental synthesis. Accordingly, we further evaluated the synthetic accessibility of the 180 identified molecules. Predicting molecular synthesizability is a complex task that typically relies on large-scale datasets, rendering it highly suitable for machine learning approaches. Several computational tools have been developed for this purpose, including Synthetic Accessibility score (SAScore) [40] and Synthetic Complexity score (SCScore) [41], both of which are widely employed in molecular and drug virtual screening. Nevertheless, these models are largely constrained by the scope of their training data and the domain expertise embedded during development, and they generally exhibit limited accuracy in predicting the synthesizability of molecules that have been successfully synthesized in experimental settings. Recently, the DeepSA model [42] was reported, which utilizes a chemistry-specific language model grounded in natural language processing algorithms to predict molecular synthesizability directly from SMILES strings. This capability enables users to prioritize compounds that are potentially more cost-effective and easier to synthesize. We assessed the synthesizability of the 180 identified molecules using the DeepSA model. As shown in **Table S2**, 79 of the SF candidates were predicted to be easy-to-synthesize (ES) based on a threshold score of 0.47, while the remaining molecules were classified as hard-to-synthesize.

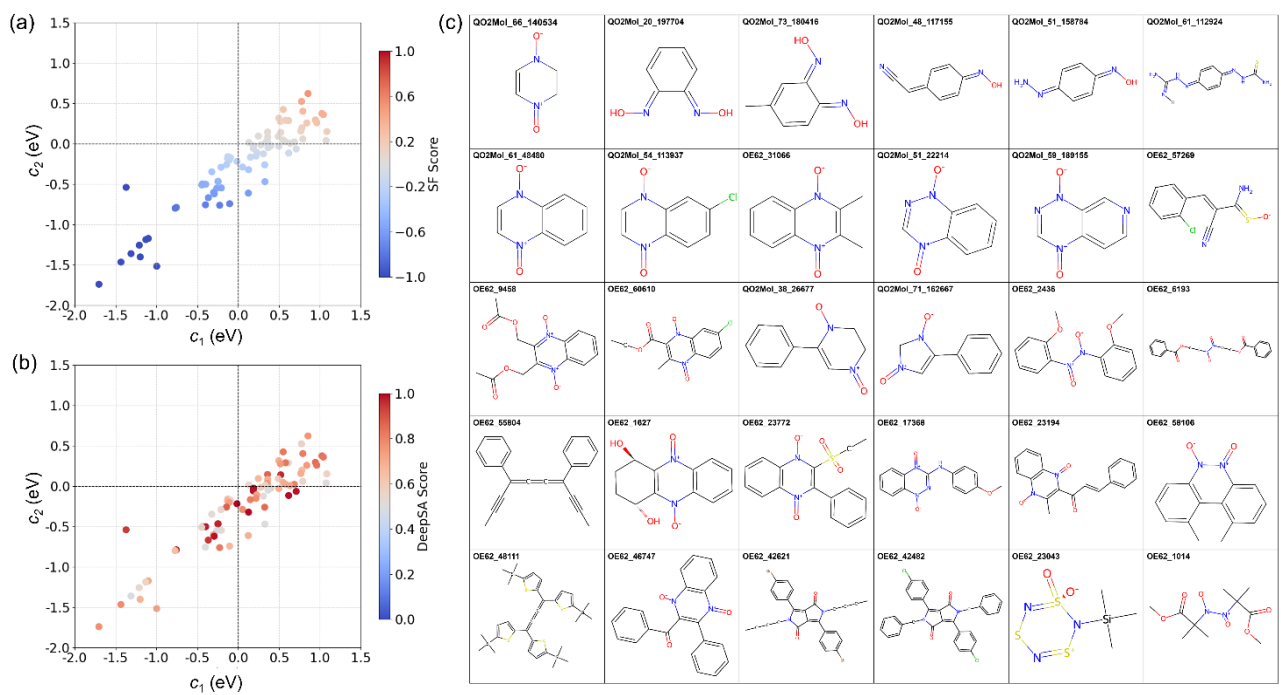

**Figure 6**. (a) SF score and (b) DeepSA score of 79 ES candidate molecules, plotted as a function of $c_1$ and $c_2$ energy according to their excitation energies calculated using the GW+BSE method. The $c_1$ and $c_2$ energy are defined as energy differences between excitation energies shown in **Figure 3b**. The colors closer to red correspond to higher SF properties and synthetic accessibility, respectively. (c) Planar diagram of 30 experimentally accessible SF molecules with positive SF scores from both TDDFT and GW+BSE calculations.

To achieve more accurate results and provide reliable guidance for experimental synthesis, the 79 ES molecules were further examined using the GW+BSE approach to compute their excitation energies. Overall, TDDFT tends to underestimate $E_{T1}$, and all candidate molecules exhibit $E_{T1}$ above 1.1 eV, corresponding to positive $c_3$ values at GW+BSE level, as illustrated in **Figure S3**. Thirty molecules display exhibit positive SF scores in both TDDFT and GW+BSE calculations, located in the upper-right region ($c_1$ and $c_2 \geqslant 0$) of **Figure 6a**. Their synthetic accessibility scores, primarily determined by molecular structure, are presented in **Figure 6b**, where colors closer to red indicate higher synthetic accessibility. The planar diagrams of these 30 organic molecular structures with positive SF scores from both TDDFT and GW+BSE calculations are plotted in **Figure 6c**. These molecules span a diverse structural space, with the majority featuring cyclic frameworks—including single benzene rings and polycyclic aromatic systems—with ring counts ranging from zero to six. The set also encompasses molecules with varied elemental compositions, including heteroatom-doped and functionalized derivatives beyond C, H, and O, such as B-doped, N-doped, and Cl-functionalized systems. Collectively, they constitute a well-rounded subset of candidate structures suitable for experimental synthesis and further investigation of SF properties.

## 4. Conclusion

To summarize, this work presents an efficient and generalizable framework for the large-scale discovery of organic SF molecules. By integrating a graph neural network trained on the FORMED database with multi-level physical validation, the framework achieves high predictive accuracy while significantly reducing computational costs. The workflow not only identifies a set of SF molecules with confirmed excited-state properties but also incorporates synthetic accessibility analysis and GW+BSE validation, thereby establishing a more robust connection between computational prediction and experimental realization. Beyond SF systems, the hierarchical strategy introduced in this study provides a transferable paradigm for the accelerated screening of other functional molecules or materials governed by excited-state phenomena, including fluorescence, phosphorescence, triplet-triplet annihilation, thermally activated delayed fluorescence.

## Data availability statement

The GNN model and structure file of SF molecules datasets that supports the findings of this study are openly available in GitHub: https://github.com/fuli-phy/SF_mols. The details of GNN model and the excitation energies calculated at TDDFT and GW+BSE level of 79 ES molecules are given in Supporting Information.

## Acknowledgements

This work was supported by the National Natural Science Foundation of China (12534012, 12474221, 12474168).

**Efficient Screening of Organic Singlet Fission Molecules Using Graph Neural Networks**

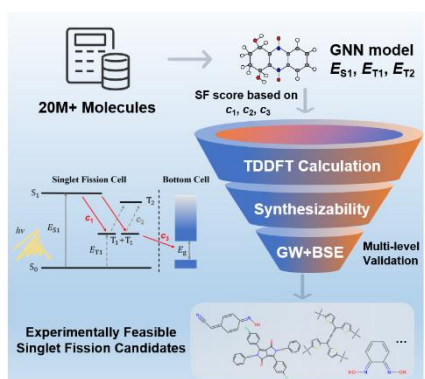

**TOC.** A high-throughput screening framework based on graph neural networks (GNNs) and multi-level validation facilitates the identification of singlet fission (SF) candidates. By efficiently predicting excitation energies across 20 million molecules, and integrating TDDFT calculations, synthetic accessibility assessments, and GW+BSE calculations, this approach yields a database of experimentally viable SF materials.